\documentclass[11pt]{article}
\usepackage{graphicx}
\usepackage{dcolumn}   
\usepackage{bm}        
\usepackage{amssymb}   
\usepackage{pslatex}

\title{A microscopic approach to the nonlinear elasticity of compressed emulsions}

\author{Ivane Jorjadze, Lea-Laetitia Pontani, Jasna Bruji\'c\\
\small Center for Soft Matter Research, Physics Department, New York University\\
\small New York University, 4 Washington Place, New York, NY 10003\\}

\begin{document}
\maketitle

\begin{abstract}
Using confocal microscopy, we measure the packing geometry and interdroplet forces as a function of the osmotic pressure in a 3D emulsion system. We find that the nonlinear elastic response of the pressure with density is not a result of the anharmonicity in the interaction potential, but of the corrections to the scaling laws of the microstructure away from the critical point. The bulk modulus depends on the excess contacts created under compression, which leads to the correction exponent $\alpha=1.5$. Microscopically, the nonlinearities manifest themselves as a narrowing of the distribution of the pressure per particle as a function of the global pressure.
\end{abstract}

The elasticity of jammed sphere packings sheds light on the broader question of how amorphous solids support stress~\cite{hecke2010,liu2010a}. In the limit of purely repulsive frictionless particles, applicable to emulsions and foams, theoretical arguments suggest a singularity at the critical random close packing density $\phi_{\textrm{c}}$~\cite{wyartT2,wyart2005}. This singularity gives rise to the surprising scaling behavior of the elastic moduli and the microstructure, as observed in numerical simulations~\cite{durian1995, makse2000, ohern2002, ohern2003, ellenbroek2009}.  In particular, the excess particle contacts created upon compression are thought to control the singularity in elasticity. However, an experimental test of the scaling law between the coordination number and the applied pressure is lacking in the literature. This is because it is experimentally difficult to access the regime very close to the jamming transition with enough resolution to test the scaling laws. Moreover, the contact network in 3D is typically hidden from view. On the other hand, bulk measurements have shown that compressed emulsions exhibit deviations from the predicted linear scaling of the osmotic pressure with the global density above random close packing~\cite{mason1995}. This scaling was conjectured to arise from the contact number dependent anharmonicity observed in the potential of a single compressed droplet~\cite{lacasse1996}. While a stiffer potential could lead to nonlinear elasticity in the limit of marginal rigidity~\cite{wyart2005}, the making and breaking of contacts are other plausible sources for the deviation~\cite{schreck2011}. A microscopic experimental approach is therefore needed to decipher the origin of the macroscopic elastic response of the system.

\begin{figure}[b!]
\begin{center}
\includegraphics[scale=0.6]{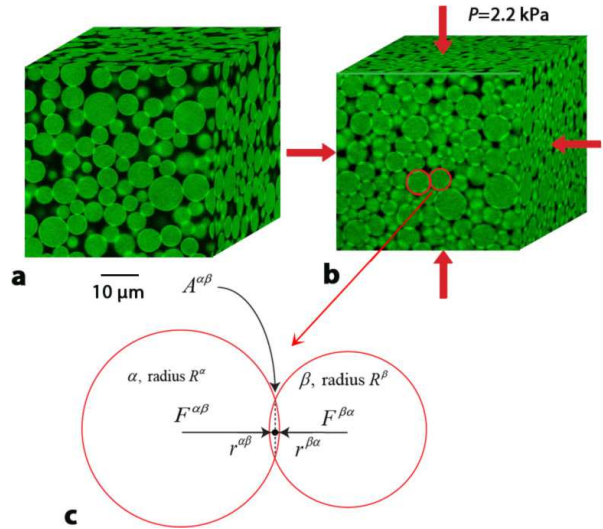}
\end{center}
\caption{\label{fig1} Confocal 3D images for emulsions creamed under gravity in (a) and compressed at $P=2.2$kPa by uniaxial centrifugation in (b). (c) A zoom of two reconstructed droplets labeled $\alpha$ and $\beta$ shows their geometric overlap area $A^{\alpha \beta}$, which is used to estimate the interdroplet force.}
\end{figure}

\begin{figure*}[t!]
\includegraphics[scale=0.5]{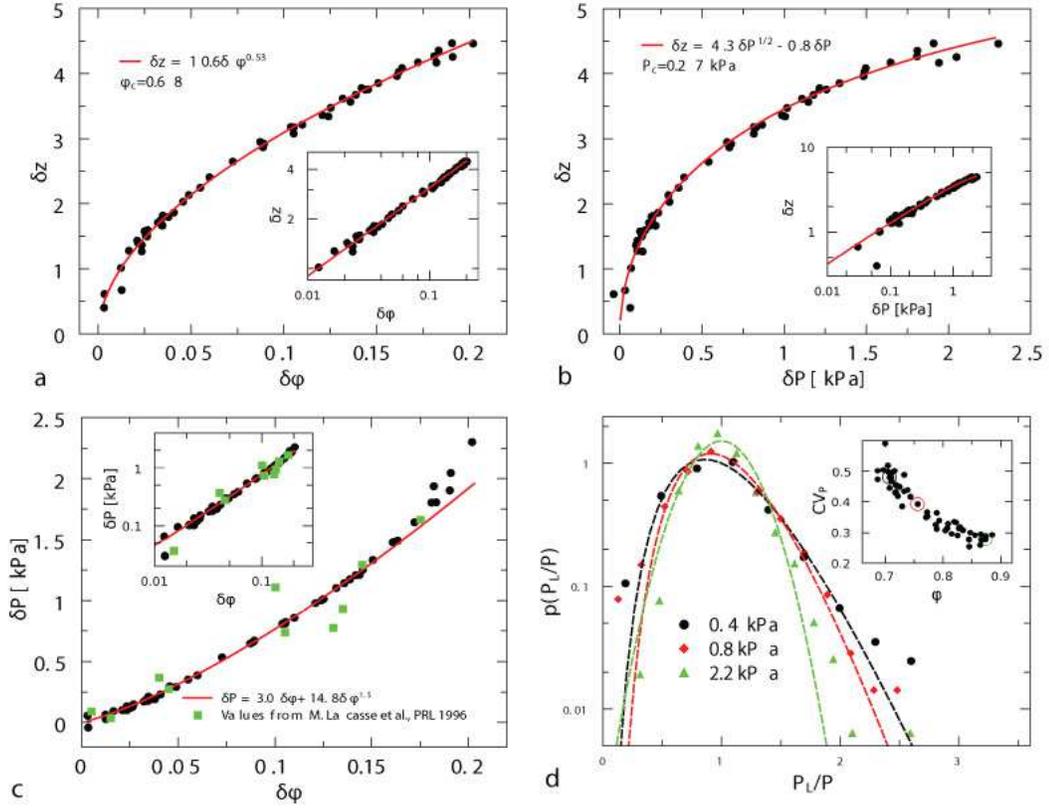}
\caption{\label{fig2} Scaling laws for the excess average number of contacts $\delta{z}$ with the change in global density $\delta \phi=(\phi-\phi_{\textrm{c}})$ in (a) and the applied pressure $\delta{P}=(P-P_{\textrm{c}})$ in (b).  A parametric plot of $\delta{P}$ versus $\delta \phi$ in (c) implies the nonlinear scaling law, in agreement with previous bulk measurements. The power-law fits are shown on log-log scales in the insets. (d) Probability density distributions of the pressure per particle $P_{\textrm{L}}$ rescaled by the mean as a function of the applied pressure. The inset shows the corresponding coefficients of variation, CV for all measured distributions. The pressure fluctuations are best fit with exponential tails at low pressures and a Gaussian distribution at high pressure. }
\end{figure*}

Here we confocally image compressed emulsions to measure the packing geometry and the interdroplet forces in 3D, from which we then infer the scaling laws between the applied pressure $P$, the average number of contacts $z$, and the global density $\phi$ away from the critical point. The statistical fluctuations of these quantities within the packing give us additional insight into the mechanisms of stress transmission in this amorphous medium. The oil-in-water emulsion system described in references~\cite{mason1995, jasna2003} is athermal with an average droplet radius $\langle R \rangle= 2.5 \textrm{ } \mu \textrm{m}$ and a polydispersity of $25\%$ preventing crystallization~\cite{chai1998}. The refractive index matched emulsion is transparent and the Nile Red dye reveals the 3D packing of droplets using a confocal microscope (Leica TCS SP5 II). The lowest compression rate is achieved by creaming under gravity, which gives a packing fraction of $\phi_{\textrm{c}} = 0.68\pm 0.02$, while centrifugation for 20 minutes at an acceleration rate of $3000g$ leads to foam-like structures with $\phi=0.88\pm 0.02$, as shown in Figs. ~\ref{fig1}A,B, respectively. Allowing the emulsion to relax to its uncompressed state over a period of several days probes a broad range of intermediate packing densities. The packings are analyzed using a Fourier transform algorithm to reconstruct the packing~\cite{jasna2003} and a navigation map tessellation to identify which particle resides in each cell volume~\cite{medvedev1995, clus2009, jorj2011}.

For each particle, we next obtain the local volume fraction $\phi_{\textrm{L}}$, the number of contacts $z_{\textrm{L}}$ and the pressure on that particle $P_{\textrm{L}}$. The statistics of the local parameters in a given packing yield the corresponding global parameters. The local volume fraction is defined as the ratio of the volume of the particle to the volume of its navigation map cell, while the global packing fraction is the total volume of the droplets divided by the volume of the box. To determine the contact network, we test whether there is a geometric overlap between the reconstructed spheres \cite{jasna2007}, as shown in Fig. ~\ref{fig1}C. This procedure leads to an estimate of the mean contact number $z\pm 0.3$, where the uncertainty is due to the resolution limit of the microscope. Moreover, the area of overlap, $A^{\alpha \beta}$, is proportional to the repulsive interdroplet force $F^{\alpha \beta}$, which is given by the Princen model~\cite{princ1983}:
\begin{equation}
F^{\alpha \beta}=\frac{\sigma}{\widetilde{R}^{\alpha \beta}} A^{\alpha \beta};
\label{eq2} 
\end{equation}
where the weighted radius $\widetilde{R}^{\alpha \beta}=\frac{2R^{\alpha}R^{\beta}}{R^{\alpha}+R^{\beta}}$ and the interfacial tension $\sigma = 9.2 \textrm{ } \textrm{mN/m}$. For small deformations, this model gives an essentially linear force law. This result is corroborated by the fact that the geometric overlap $A^{\alpha \beta}$ agrees with the area of highlighted fluorescence intensity between the droplets, i.e. the area of deformation~\cite{jasnaT}. Moreover, minimizing the energy of this system of linear springs to satisfy force balance does not move any of the particles beyond the resolution of the technique, which is one third of the voxel size, i.e. 100nm. Note that the incompressibility of the emulsion leads to a divergence of the pressure at very high compression. In addition, the ability to lose or create contacts is another source of nonlinearity~\cite{schreck2011}. 

\begin{figure*}[t!]
\includegraphics[scale=0.5]{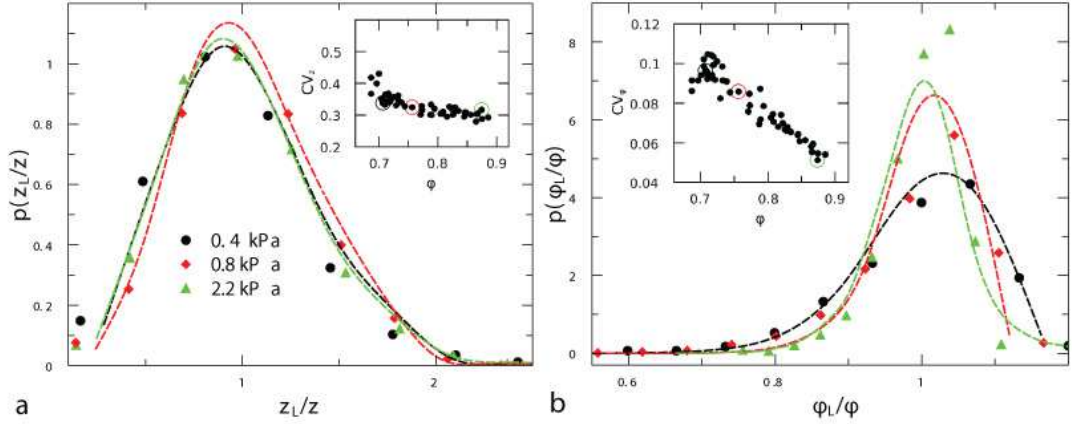}
\caption{\label{fig4} Probability density distributions of local parameters $z_{\textrm{L}}$ in (a) and $\phi_{\textrm{L}}$ in (b) are rescaled by the mean for three different global pressure values. Insets show the corresponding coefficients of variation, CV, for all measured distributions. The data are well fit with the granocentric model. }
\end{figure*}

The measurement of the forces then allows us to calculate the local pressure per particle $P_{\textrm{L}}$, which is equal to the trace of the corresponding Cauchy stress tensor, $s^{\alpha}$:
\begin{equation}
s_{ij}^{\alpha}=\frac{1}{6V^{\alpha}}  \sum_{\beta} \left( F_{i}^{\alpha \beta}r_{j}^{\alpha \beta} + F_{j}^{\alpha \beta}r_{i}^{\alpha \beta}  \right);
\label{eq1} \end{equation}
where indices $i, j$ represent vector projections of each contact force $F^{\alpha \beta}$, $V^{\alpha}$ is the volume of the navigation cell of the droplet $\alpha$, $r^{\alpha \beta}$ is a vector from the center of $\alpha$ to its point of contact with the droplet $\beta$, and the sum of the force moment tensor is taken over all contacting droplets.  Applying Eq. (2) to the entire packing gives the global pressure $P$. We find that the diagonal elements of the global stress tensor are close to each other in magnitude, suggesting that the compression of the emulsion is isotropic due to the absence of friction. We repeat these measurements on $50$ samples with global pressure values ranging from $0.27\pm0.05\textrm{ kPa}$ under gravity to $2.29\pm0.05 \textrm{ kPa}$.

Let us first assess how the excess contacts grow with the global density. Our system of frictionless spheres reaches isostatic equilibrium at the critical point, which fixes its average coordination number to $z_{\textrm{c}}=2d$, where $d=3$ is the dimension~\cite{alexander1998, wyart2005}. To deduce the volume fraction we take into account the shape deformation of each droplet away from its spherical shape, such that the droplet volume cannot be larger than that of its cell. In Fig. 2A we show the scaling of excess contacts with the square root of the packing density $\delta{z}=z-z_{\textrm{c}}= z_0\sqrt{(\phi-\phi_{\textrm{c}})}=z_0\sqrt{\delta \phi}$ over the experimental range. This geometric result is in good agreement with previous numerical simulations and theoretical arguments~\cite{durian1995, ohern2003, wyart2005}, as well as 2D data on frictional particles~\cite{behring2007} and foams~\cite{katgert2010}. Moreover, the value of the prefactor $z_0=10.6$ is in good agreement with the one obtained from numerical simulations of bidisperse spheres close to the jamming point, $z_0=8.4\pm0.5$~\cite{ohern2003}. This fit determines a higher value of $\phi_{\textrm{c}}=0.68$ for the polydisperse packing than the predicted monodisperse random close packing density, $\phi_{\textrm{c}}=0.64$~\cite{makse2008}. The obtained value is in excellent agreement with the density measured at the lowest pressure. 

By contrast, the scaling of excess contacts with the applied pressure in {Fig. 2B} cannot be fit by the square root law predicted for a network of harmonic spheres close to the jamming point~\cite{wyart2005}: $\delta{z} \sim \sqrt{P-Pc}$, where $P_c=0.27$kPa is the residual pressure under gravity. This scaling behavior was observed earlier in simulations of harmonic, frictionless particles \cite{durian1995, ohern2003}. Surprisingly, such a scaling was observed experimentally in the case of frictional disks over a narrow range of $1\%$ in density~\cite{behring2007}. Our experiments are the first to test this relation in frictionless packings and in 3D, over a much broader range of packing densities up to $0.2$ above $\phi_{\textrm{c}}$. The data shows important deviations from the scaling prediction, explained below.
 
Combining the two square root laws for the scaling of excess contacts with density and pressure predicts a linear parametric relationship between pressure and density, where $\delta{P}=P-P_c=P_0 {\delta{\phi}}^{\gamma-1} $, $P_0$ is the prefactor and $\gamma=2$ is the exponent in the harmonic interaction potential~\cite{ohern2003,wyartT2}. However, the deviations of the data from the square root law in {Fig.~2B} lead to the nonlinear relationship of pressure versus density presented in Fig. 2C. These measurements are in good agreement with those obtained from bulk experiments~\cite{mason1995}, which measure the pressure macroscopically. The agreement between the two data sets gives additional validity to the harmonic assumption in Eq. (1) used to calculate the pressure in our system. Therefore, the observed deviations from the linear scaling are likely to come form the formation of contacts upon compression and not the proposed anharmonicity in the potential~\cite{lacasse1996}.

Theoretically, these rearrangements can be understood in terms of the corrections to scaling away from the critical point.
It has been shown that for generic random elastic networks, all the elastic moduli must vanish linearly with the excess coordination $G\sim B\sim \delta z$~\cite{wyartT2, wyartT3}, as is observed numerically~\cite{ellenbroek2009}. However, packings of particles differ from random networks in the following sense: their geometry is constrained by the fact that all the contact forces between the particles are positive. It was shown that this constraint implies that the bulk modulus must have an additional contribution that does not vanish at jamming, that is $B\equiv\phi \partial P/\partial \phi\approx C_1+ C_2 \delta z$, where $C_1$ and$C_2$ are constants. Moreover, the parameter $\delta z$ has been shown to govern the crossover from the isostatic behavior close to jamming to the continuum behavior at the large scale~\cite{ellenbroek2010}. Since the scaling relation $\delta z\sim \sqrt{\delta{\phi}}$ holds over the experimental range in Fig. 2A, we get $B \approx C_1+C_2' \sqrt {\delta \phi}$~\cite{wyartT3}. If we assume that, apart from the creation of new contacts, no plastic rearrangements occur, this equation applies for all $\phi$ and can be integrated and expanded to give the two leading order terms in $\delta \phi$,
\begin{equation}
	P-P_{\textrm{c}}=P_0(\phi-\phi_{\textrm{c}})+P_1(\phi-\phi_{\textrm{c}})^\alpha
\label{eq2} 
\end{equation}
where $P_{\textrm{c}}=0.27$~kPa is the experimental value of the pressure at the lowest compression, $P_0=3.0$~kPa and $P_1=14.8$~kPa are prefactors obtained from the fit of the data in Fig. 2C, and $\alpha=1.5$ is fixed as the universal exponent for the first correction to scaling. The data is in excellent agreement with the theory in the range up to $\delta \phi=0.18$, beyond which the incompressibility of the droplets causes a divergence in the pressure. Remarkably, converting the prefactor $P_0$ in the linear term to the units of pressure defined in numerical simulations of bidisperse packings close to jamming recovers the same value to one decimal place~\cite{ohern2003}. The success of the fit in Fig. 2C with Eq. (3) implies that the data in Fig. 2B can also be explained in terms of the corrections to the scaling of excess contacts with pressure in the form $\delta z= D_1\sqrt{\delta P}-D_2\delta P$, where the prefactors $D_1=4.3$~kPa$^{-1/2}$ and $D_2=0.8$~kPa$^{-1}$ are obtained from the fit in the figure. The prefactor values agree with the parametric prediction of the fits in Figs. 2A,C, which testifies to the accuracy of the experimental data and the validity of the model. 

Microscopically, the distributions of the pressure per particle $P_{\textrm{L}}$ for three different pressures are shown in Figure 2D. The tails of the distributions narrow down from exponential to Gaussian as a function of the applied pressure, as shown in Fig. 3C. This is consistent with theoretical considerations for the distribution of forces close to the jamming point~\cite{coppsm1996, jasna2003, wang2010} and the randomization of stress that occurs as the pressure is increased~\cite{dinsm2008}. The change in shape of the local distributions with the applied pressure contributes to the nonlinear elastic response of the material. Finally, we show the microscopic distributions of $z_{\textrm{L}}$ and $\phi_{\textrm{L}}$ as a function of pressure in Fig. 3. It is interesting to note that rescaling the contact distribution by the mean collapses the distributions at different pressures. This is also confirmed by the invariant coefficient of variation shown in the inset. By contrast, the distributions of local packing fraction do not collapse with the pressure, but become more peaked. The distributions in Figs. 3A,B are superimposed with the theoretical predictions of the granocentric model~\cite{clus2009, newh2011}, which takes the measured average number of contacts, neighbors and the global density as inputs and generates the local fluctuations in each input parameter. This model is based on two local random processes: filling the space around each particle with solid angles contributed by the neighbors and choosing some of those neighbors to be contacts to ensure rigidity. The good agreement between the data and the model suggests that random processes on the level of the first neighbor shell are sufficient in explaining the diversity of local configurations.  

In conclusion, our data show corrections to the linear scaling of the pressure with density that arise because the bulk modulus is not constant near jamming, but depends linearly on the  
excess contacts in the force network. We also show that the interaction potential is nearly harmonic, independent of the compression. Therefore, the first correction to scaling, which only takes into account the creation of contacts but not particle rearrangements or the incompressibility of the droplets, is sufficient in accounting for the nonlinear dependence over a wide range of densities. This study reveals the mechanism by which emulsions are stiff to compression and demonstrates that confocal microscopy is a powerful tool to decouple the geometric and stress-bearing elements of the anomalous scaling of the bulk modulus in compressed emulsions.  

We would like to acknowledge Matthieu Wyart for pointing out the amplitude of the corrections to scaling and Eric Vanden-Eijnden for enlightening discussions. J. B. holds a Career Award at the Scientific Interface from the Burroughs Wellcome Fund. This work was supported partially by the MRSEC Program of the National Science Foundation under Award Number DMR-0820341 and the National Science Foundation Career Award 0955621.

\bibliography{mybib}
\end {document}